\documentclass[10pt, conference, compsocconf]{IEEEtran}
\usepackage{amsmath}
\usepackage{amsfonts}
\usepackage{bm}
\usepackage{color}
\usepackage{graphicx}
\usepackage[noadjust]{cite}
\usepackage{url}
\usepackage{setspace}
\usepackage{subcaption}
\usepackage{paralist}
\setdefaultleftmargin{10pt}{}{}{}{}{}

\newcommand{\reffig}[1]{Figure \ref{#1}}
\newcommand{\reftable}[1]{Table \ref{#1}}

\newcommand{\xhdr}[1]{\vspace{0mm}\noindent{{\bf #1}}}

\usepackage{booktabs}
\usepackage{mathtools}

\title{Modeling Ambiguity, Subjectivity, and Diverging Viewpoints\\
in Opinion Question Answering Systems\vspace{-0.2in}}

\author{\IEEEauthorblockN{Mengting Wan}
\IEEEauthorblockA{University of California, San Diego\\
m5wan@eng.ucsd.edu}
\and
\IEEEauthorblockN{Julian McAuley}
\IEEEauthorblockA{University of California, San Diego\\
jmcauley@eng.ucsd.edu}}
\begin{document}
\maketitle
\begin{abstract}
Product review websites provide an incredible lens into the wide variety of opinions and experiences of different people, and play a critical role in helping users discover products that match their personal needs and preferences. To help address questions that can't easily be answered by reading others' reviews, some review websites also allow users to pose questions to the community via a question-answering (QA) system. As one would expect, just as opinions diverge among different reviewers, answers to such questions may also be subjective, opinionated, and divergent. This means that answering such questions automatically is quite different from traditional QA tasks, where it is assumed that a single `correct' answer is available. While recent work introduced the idea of question-answering using product reviews, it did not account for two aspects that we consider in this paper: (1) Questions have multiple, often divergent, answers, and this full spectrum of answers should somehow be used to train the system; and (2) What makes a `good' answer depends on the asker and the answerer, and these factors should be incorporated in order for the system to be more personalized. Here we build a new QA dataset with 800 thousand questions---and over 3.1 million answers---and show that explicitly accounting for personalization and ambiguity leads both to quantitatively better answers, but also a more nuanced view of the range of supporting, but subjective, opinions.
\end{abstract}

\section{Introduction}

User-generated reviews are a valuable resource to help people make decisions. Reviews may contain a wide range of both objective and subjective product-related information, including features of the product, evaluations of its positive and negative attributes, and various personal experiences and niche use-cases.
Although a key factor in guiding many people's decisions, 
it can be
time-consuming for a 
user to 
digest the content in large volumes of reviews, many of which may not be relevant to their own opinions or interests.

In addition to passively searching for information that users are interested in among reviews, a number of e-commerce websites, such as \textit{Amazon} and \textit{ebay}, also provide community question answering systems where users can ask and answer specific product-related questions. While such systems allow users to seek targeted information (as opposed for searching for it in reviews), asking the community is still time-consuming in the sense that the user must wait for a response, and even then may have quite different preferences from the user who answers their questions.

The above issues motivate us to study systems that help users to automatically navigate large volumes of reviews in order to locate relevant and informative opinions, in response to a particular query. 

This kind of `opinion question answering' system (opinion QA) is quite different from typical community question answering (cQA) systems. 
In particular, 
traditional cQA systems are usually concerned with \emph{objective} information, such that answers can be generated by constructing and exploring a knowledge-base which is composed of facts.

\begin{figure}
\centering
\includegraphics[width=0.5\textwidth]{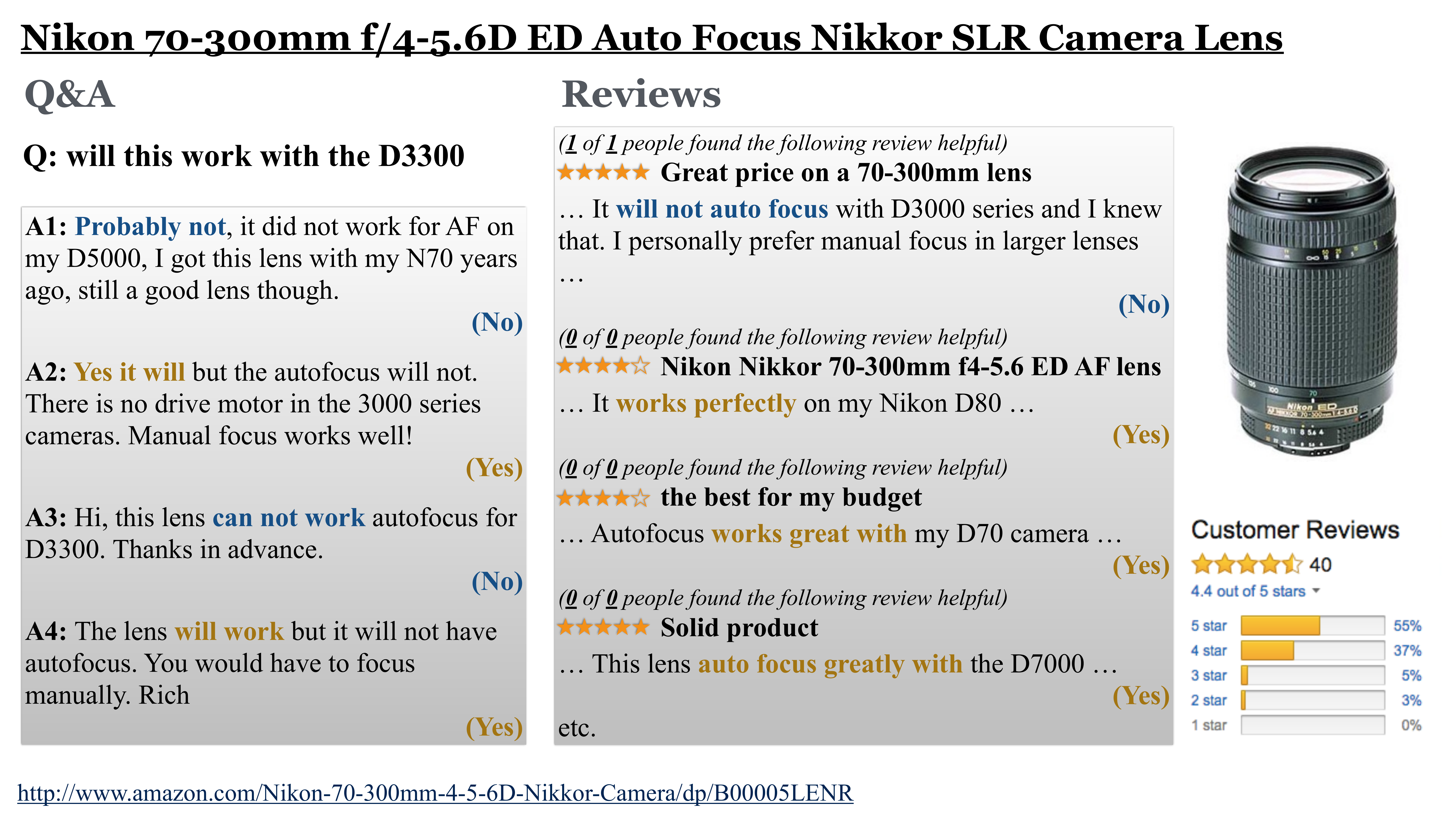}
\caption{A real opinion QA example from \emph{Amazon.com}. 
The left box shows answers provided by the community, demonstrating the divergent range of responses. The right box shows the type of system we develop to address such questions, mining divergent and subjective opinion information from product reviews.
}\label{figure:subjectiveQA}
\end{figure}

However, for the opinion QA problem, users often ask for \emph{subjective} information, such as ``Is this a good lens for my Nikon D3300 camera?'' 
Such a seemingly simple question is complex because it depends on (a) objective information (is the lens even compatible?); (b) subjective information (whether it's `good' is a matter of opinion); and (c) personalization (which answer is correct \emph{for the user asking the question}; are they an expert? an amateur? on a budget? etc.).
Perhaps not surprisingly, opinion QA systems generate a wide variety of subjective and possibly contradictory answers (see Figure \ref{figure:subjectiveQA}, from \emph{Amazon}).

Ideally answers to this kind of question should leverage data describing personal opinions and experiences, 
such as the kind of information available in product reviews. %
To build systems capable of leveraging such information, a series of methods
\cite{moghaddam2011aqa,yu2012answering,mcauley2016addressing} for product-related opinion questions answering 
have been developed. These methods can automatically retrieve potential answers from reviews based on different linguistic features. Many of these %
approaches
develop information
retrieval
systems where traditional text similarity measures are explored and text fragments are filtered based on question types or the attributes of the product that users 
refer to
in their 
questions \cite{moghaddam2011aqa,yu2012answering}. 

Recently, a supervised approach, \textit{Mixtures of Opinions for Question Answering} (MoQA) \cite{mcauley2016addressing}, was developed for opinion QA systems using product reviews.
There,
product-related questions were categorized into two types as follows:

\begin{itemize}
\item \textbf{Binary questions.} 
A large fraction
of questions in real-world opinion QA data are binary questions where answers amount to either `Yes' or `No'. 
Such answers can easily be detected (i.e., to build a labeled dataset) in a supervised setting using a binary classifier \cite{yesno}.
When addressing binary questions, we are interested both in mining relevant opinions from reviews, but also providing a yes/no answer directly.
\item \textbf{Open-ended questions.} In addition to binary questions, a significant number of product-related questions are open-ended, or compound questions (etc.). It is usually impractical to answer such questions directly with an automated system.
Instead, we are more interested in learning a good relevance function which can help us retrieve useful information from reviews,
so that the user can be aided in reaching a conclusion themselves. 
\end{itemize}

In this paper, we continue to study these two types of questions.
Where we extend existing work, and the main contribution of our paper, is to explicitly account for the fact that questions may have multiple, subjective, and possibly contradictory answers.\footnote{Note that even binary questions may still be subjective, such that both `yes' and `no' answers may be possible.} We evaluate our system by collecting a new QA dataset from \emph{Amazon.com}---consisting of 800 thousand questions and 3.1 million answers, which uses \emph{all} of the available answers for training (in contrast to previous approaches, where each question was associated with only a single answer).

Our main goals are to show quantitatively that by leveraging multiple answers in a supervised framework we can provide more accurate responses to both subjective and objective questions (where `accurate' for a subjective question means that we can correctly estimate the distribution of views). Qualitatively, we aim to build systems that are capable of presenting users with a more nuanced selection of supporting evidence, capturing the full spectrum of relevant opinions.

\subsection{Ambiguity and Subjectivity in Opinion QA Systems}

Addressing this new view of question-answering is challenging, and requires new techniques to be developed in order to make use of multiple, possibly contradictory labels within a supervised framework.
We identify two main perspectives from which ambiguity and subjectivity in product-related opinion QA systems can be studied:
\begin{itemize}
\item \textbf{Multiple Answers.} We notice that in previous studies, only one ground-truth
answer is included for each
question.
However, in real-world opinion QA systems, multiple answers are often available.
We find this to be true both for binary and open-ended questions.
When multiple answers are available,
they often describe different aspects of the questions or different personal experiences. 
By including
multiple answers 
at training time,
we expect that the relevant reviews retrieved
by
the system
at test time
should cover those subjective responses more comprehensively.

\item \textbf{Subjective Reviews.} In addition, as indicated in traditional opinion mining studies, reviews as reflections of users' opinions may be subjective since different reviewers may have different expertise and bias. In some review websites, such as \emph{Amazon.com}, review rating scores and review helpfulness can be obtained, which could be good features reflecting the subjectivity of the reviews.
Intuitively, subjective information may affect the
language
that users apply to express their opinion so that their reviews should be handled to address questions accordingly. For example, `picky' reviewers may %
tend to
provide negative responses while `generous' reviewers may usually provide
more favorable
information about the product. This motivates us to apply user modeling approaches and incorporate more subjective review-related features into opinion QA systems. 
\end{itemize}

The
above observations provide us with a strong motivation to study
ambiguity and subjectivity from the perspective of multiple answers and subjective reviews in opinion QA systems. We conclude by stating the problem specifically as follows:
\begin{quote}
\textbf{Goal:} \emph{Given a question related to a product, we would like to determine how relevant each review of that product is to the question with emphasis on modeling \textbf{ambiguity} and \textbf{subjectivity}, where `relevance' is measured in terms of how helpful the review will be in terms of identifying the proper response (or responses) to the question.}
\end{quote}

\subsection{Contributions}

To our knowledge, our study
is 
the
first one to systematically model 
ambiguity and subjectivity in opinion QA systems. 
We provide a new dataset 
consisting of
135 thousands product from \emph{Amazon}, 808 thousand questions, 3 million answers and 11 million reviews.\footnote{Data and code are available on the first author's webpage.}
By modeling ambiguity
in
product-related questions, this study
not only
bridges QA systems and reviews, but also bridges opinion mining and the idea of `learning from crowds.' For both binary and open-ended questions, we successfully develop a model to handle
multiple (and possibly conflicting)
answers
and incorporate subjective features, 
where labels are predicted for binary questions,
and a relevance-ranked list of reviews is surfaced to the user. Quantitatively, we show that modeling ambiguity and subjectivity leads to substantial performance gains in terms of the accuracy of our question answering system.

\section{Background}
In this study, we
build upon
the mixture of experts (MoE) framework
as used previously by \cite{mcauley2016addressing}.
We enhance this approach by
modeling
ambiguity and subjectivity from the perspectives of answers and reviews.
Before
introducing the complete model, we introduce standard relevance measures and the mixture of experts (MoE) framework as background knowledge.
The basic notation 
used throughout this paper is
provided in \reftable{table:notation}. 

\subsection{Standard Relevance Measures}
We first describe two kinds of similarity measures for relevance ranking in the context of our opinion QA problem as follows.

\subsubsection{Okapi BM25} 
One of the standard
relevance ranking measures for information retrieval,  Okapi BM25
is a bag-of-words
`tf-idf'-based ranking function 
that
has been successfully applied in a number of
problems
including QA tasks \cite{manning2008introduction,jones2000probabilistic}. Particularly, for a given question $q$ and a review $r$, the standard BM25 measure is defined as
\begin{equation}
\small
\mathit{bm25}(q,r)=\sum_{i=1}^n \frac{\mathit{idf}(q_i) \times f(q_i,r)\times(k_1+1)}{f(q_i,r)+k_1\times(1-b+b\times\frac{|r|}{\mathit{avgrl}})},
\end{equation}
where $q_i,i=1,\ldots,n$
are keywords in $q$, $f(q_i,r)$ denotes the frequency of $q_i$ in $r$, $|r|$ is the length of review $r$ and $\mathit{avgrl}$ is the average review length among all 
reviews.%
\footnote{In practice
we set $k_1=1.5$ and $b=0.75$.} Here $\mathit{idf}(q_i)$, 
the
inverse document frequency of $q_i$, is defined as
\begin{equation}
\small
idf(q_i)=\log\frac{N-n(q_i)+0.5}{n(q_i)+0.5},
\end{equation}
where $N=|\mathcal{R}|$ is the total number of reviews and $n(q_i)$ is the number of reviews which contain $q_i$.

\subsubsection{Rouge-L} 
Next we consider
another similarity measure, Rouge-L \cite{lin2004rouge}, which is a Longest Common Subsequence (LCS) based statistic.
For a question $q$ and a review $r$, if the length of their longest common subsequence
is denoted as $\mathit{LCS}(q,r)$, then we have
$R_{\mathit{LCS}}=\mathit{LCS}(q,r)/|q|$ and $P_{\mathit{LCS}}=\mathit{LCS}(q,r)/|r|$. Now Rouge-L is defined as
\begin{equation}
\small
F_{\mathit{LCS}} = \frac{(1+\beta^2)R_{\mathit{LCS}}P_{\mathit{LCS}}}{R_{\mathit{LCS}}+\beta^2P_{\mathit{LCS}}},
\end{equation}
where $\beta = P_{\mathit{LCS}}/R_{\mathit{LCS}}$.

\subsection{Mixtures of Experts}
\label{sec:moeBackground}

Mixtures of experts (\textbf{MoE}) \cite{jordan1994hierarchical} is 
a supervised learning approach that smoothly combines 
the outputs of several `weak' 
classifiers
in order to generate predictions.
Here, this method can be applied for opinion QA systems %
where each individual review is regarded as a weak classifier that makes a prediction about the response to a query. For each classifier (review), we output a relevance/confidence score (how relevant is this review to the query?), as well as a 
prediction (e.g.~is the response `yes' based on the evidence in this review?).
Then an overall prediction can be obtained for a particular question by combining outputs from all reviews of a product, weighted by their confidence. 

\subsubsection{MoE for binary questions}
For a binary question, each classifier 
produces
a probability associated with a positive label, i.e., a probability that the answer is `yes.' Suppose for a question $q$, 
the associated features (including the text itself, the identity of the querier, etc.)
are denoted
$X_q$ and the label for this question is denoted as $y_q$ ($y_q\in\{0,1\}$). Then we have
\begin{equation}
\small
P(y_q|X_q) = \sum_{r \in \mathcal{R}_q} \overbrace{P(r|X_q)}^{\text{how relevant is } r} \times\overbrace{P(y_q|r, X_q)}^{\text{prediction from } r}, \label{eq:MoE_binary}
\end{equation}
where $r$ is a review among the set of reviews $\mathcal{R}_q$ 
associated with the question $q$.
In \eqref{eq:MoE_binary}, $P(r|X_q)$ measures the confidence of review $r$'s ability in terms of responding 
to the
question $q$, and $P(y_q|r, X_q)$ is the prediction for $q$ given by review $r$. 
These 
two terms can be modeled as follows:
\begin{equation}
\small
\begin{aligned}
\text{(Relevance) }\quad & P(r|X_q) = \exp(v_{q,r})/\sum_{r'\in \mathcal{R}_q}\exp(v_{q,r'}); \\
\text{(Prediction) }\quad & P(y_q=1|r, X_q) = \sigma (w_{q,r}) \label{eq:rel_pred},
\end{aligned}
\end{equation}
where $\sigma(x)=1/(1+\exp(-x))$ is the sigmoid function. 
Here $v_{q,r}$ and $w_{q,r}$ are real-valued (i.e., unnormalized) `relevance' and `prediction'
scores
where multiple question and review related features can be involved.

\subsubsection{MoE for open-ended questions}
\label{sec:moeOpenEnded}

Similarly, for an open-ended question, we may be interested in whether a `true' answer $a_q$ is preferred over some arbitrary non-answer $\bar{a}$.
For this
we have a similar MoE structure as follows:
\begin{equation}
\small
P(a_q > \bar{a}| X_q) = \sum_{r\in\mathcal{R}_q} P(r|X_q) P(a_q>\bar{a}|r, X_q). \label{eq:MoE_open}
\end{equation}
The relevance term can be kept 
the same while we have a slightly different prediction term:
\begin{equation}
\small
P(a_q>\bar{a}|r,X_q) = \sigma(w_{a_q>\bar{a},r}).
\end{equation}
Here $w_{a_q>\bar{a},r}$ is a real-valued `prediction' score where multiple answer and review 
features can be included.

\begin{table}
\small
\begin{tabular}{p{0.59in}p{2.5in}}
\toprule
Notation & Description \\
\midrule
$q$, $\mathcal{Q}$ & question, question set \\
$a$, $\mathcal{A}_q$ & answer, answer set to question $q$ \\
$r$, $\mathcal{R}_q$ & review, review set 
associated with
question $q$ \\
$y_q\in\{0,1\}$ & label for a binary question $q$ \\
$p_q $ & probability of assigning a positive %
label to 
$q$\\
$a_q > \bar{a}$ & answer $a_q$ is preferred over an alternative $\bar{a}$ for $q$\\
$p_{a_q}$ & probability of answer $a_q$ being preferred over $\bar{a}$ for $q$ \\
$v_{\cdot,r},w_{\cdot,r}$ & `relevance' and `prediction' scores\\
$\bm{f}_q, \bm{f}_a, \bm{f}_r$ & unigram text features of $q$, $a$ and $r$\\
$\bm{s}(q,r)$ & pre-computed similarities between $q$ and $r$\\
$y_{q,j}$ & the $j$-th label provided for a binary question $q$ \\
$n_q^+, n_q^-, n_q$ & numbers of positive, negative and total provided labels for a binary question $q$\\
$r_q$ & $r_q=n_q^+/n_q$, the fraction of positive labels for $q$\\
$\alpha_q, \beta_q$ & ``sensitivity'' and ``specificity'' regarding $q$ \\
$\bm{h}_r$ & helpfulness features for review $r$ \\
$u_r, e_{u_r}, b_{u_r}$ & reviewer who provides review $r$, expertise of reviewer $u_r$, bias of reviewer $u_r$ \\
$rt_r$ & rating score associating review $r$ \\
\bottomrule
\end{tabular}\vspace{0.05in}
\caption{Basic notation in this study.}\label{table:notation}
\end{table}

\subsection{Relevance and Prediction with Text-Only Features}

As described above,
for a binary question, the probability associated with
a
positive (i.e., `yes') label $P(y_q=1|X_q)$ ($p_q$ in shorthand) can be modeled using an {MoE} framework where each review
is
regarded as a weak classifier. If only one label 
is
included for a question in the training procedure, we can 
train by
maximizing the following log-likelihood:
\begin{equation}
\small
\begin{aligned}
    \mathcal{L} = \log P(\mathcal{Y} | \mathcal{X})
    = \sum_q [y_q \log p_q + (1-y_q) \log(1-p_q)] \label{eq:log-lik}
\end{aligned}
\end{equation}
where $\Theta$ includes all
parameters and $p_q$ is modeled
as in \eqref{eq:MoE_binary}. 

A number of 
features can be applied
to define the
`relevance' ($v_{q,r}$) and `prediction' ($w_{q,r}$) functions. Previously in \cite{mcauley2016addressing}, only text features were used
to define
pairwise similarity measures and bilinear models.
Starting with the same text-only model,
suppose $\bm{f}_q$ and $\bm{f}_r$ are vectors with length $N$
that
represent bag-of-words text features for question $q$ and review $r$.
Then we
define
the `relevance' function as follows:
\begin{equation}
\small
v_{q,r} = \overbrace{\left< \bm{\kappa}, \bm{s}(q,r)\right>}^{\text{pairwise similarities (bm25 etc.)}} + \overbrace{\left< \bm{\eta}, \bm{f}_q \circ \bm{f}_r \right>}^{\text{term-to-term similarity}}, \label{eq:score_basic_v}
\end{equation}
where $\bm{x} \circ \bm{y}$ is the Hadamard product.
Note that
we have two parts in
$v_{q,r}$: (1) a weighted combination of
state-of-the-art
pairwise similarities;
and (2) a parameterized term-to-term similarity. Following \cite{mcauley2016addressing}, we include
BM25 \cite{manning2008introduction} and Rouge-L \cite{lin2004rouge} measures
in $\bm{s}(q,r)$. Recall that
the purpose of this function
is to learn a set of parameters $\{\bm{\kappa}, \bm{\eta}\}$
that ranks reviews in order of relevance.

In addition, we define the following prediction function:
\begin{equation}
w_{q,r} = \overbrace{\left< \bm{\mu}, \bm{f}_q \circ \bm{f}_r \right>}^{\text{interaction between q.~\& r.~text}} + \overbrace{\left<\bm{\xi}, \bm{f}_r\right>}^{\text{prediction from r. text}} \label{eq:score_basic_w}.
\end{equation}
The
idea here is that the first term
models the interaction between
the
question and review
text, 
while the second models only the review (which can capture e.g.~sentiment words in the review).
\paragraph*{Training}
Finally, to optimize the parameters
$\Theta=\{\bm{\kappa}, \bm{\eta}, \bm{\mu}, \bm{u}\}$ from \eqref{eq:score_basic_v} and \eqref{eq:score_basic_w}, we apply L-BFGS \cite{nocedal2006numerical}.
To avoid overfitting, this model also includes a simple $L_2$ 
regularizer on all model parameters. 
\section{Ambiguity and Subjectivity in Binary Questions}

So far, we have followed the basic approach of \cite{mcauley2016addressing}, assuming text-only features, and a single label (answer) associated with each question.
But as we find in our data (see Section \ref{sec:dataset}), 
responses in real-world opinion QA systems have significant ambiguity, even for binary questions.
However, in previous 
studies, only a single response was considered to each question. 
In this section we develop additional machinery allowing us to model ambiguity and subjectivity, and in particular to handle training data with multiple (and possibly contradictory) answers.

\subsection{Modeling
Ambiguity: Learning with Multiple Labels.} 
\label{sec:klem}

Notice that 
in the
previous log-likelihood
exppression
\eqref{eq:log-lik}, only one label can be included for each question. Below are two options to extend this framework to handle multiple labels.  %

\subsubsection{\textbf{KL-MoE}} A straightforward approach is 
to
replace the single label $y_q$ in \eqref{eq:log-lik} by the the fraction of positive labels $r_q = n_q^+/(n_q^++n_q^-)$, where $n_q^+, n_q^-$ are the number of positive and negative (yes/no) answers for question $q$. If we assume that for a question $q$, the response provided from 
the
answers
given
follows $\mathit{Bernoulli}(r_q)$ and the response predicted from reviews follows $\mathit{Bernoulli}(p_q)$, then the objective function
\begin{equation}
\sum_q[r_q\log p_q + (1-r_q)\log(1-p_q)] \label{eq:log-lik-KL}
\end{equation}
can be regarded as the summation of the KL-divergences between answers and predictions for all questions.

\subsubsection{\textbf{EM-MoE}}
Note
that only the 
\emph{ratio} of positive and negative labels is included in
the
previous KL-divergence
loss
\eqref{eq:log-lik-KL}, while the real counts of positive and negative labels are discarded. However, this fraction may not be enough to model the strength of the ambiguity (or controversy) 
in
the question. For example, a question with 10 positive 
and 10 negative labels seems more controversial than a question with 1 positive
and 1 negative label. However, their positive/negative 
ratios $r_q$
are the same. 

To distinguish such cases,
instead of applying
a
fixed
ratio
$r_q$, we use two sets of parameters, allowing us to
incorporate
multiple noisy labels at training time, 
and to update (our noisy estimate of) $r_q$ based on
multiple
labels $y_{q,j}$ and generated predictions $p_q$ iteratively using 
the
EM-algorithm. 

Specifically, for a binary question $q$, %
we model its
`true'
answer $y_q$
as an unknown with
probability distribution $P(y_q = 1 | X_q, \Theta)$,
which is assumed to generate the
provided (noisy) labels $y_{q,j}$  ($j=1,\ldots,n_q$)
independently.
Then
the joint probability of
the observed
labels is given 
by
\begin{equation}
\small
\begin{aligned}
& P(y_{q,1},
\ldots
,y_{q,n_q} | X_q, \Theta) \\
=& \sum_{i\in\{0,1\}} P(y_{q,1}, \ldots, y_{q,n_q} | y_q=i, X_q, \Theta) P(y_q=i | X_q, \Theta)\\ 
=& \sum_{i\in\{0,1\}} \left( \prod_{j=1}^{n_q} P(y_{q,j}|y_q=i, X_q, \Theta) \right) P(y_q=i | X_q, \Theta)\label{eq:multi}
\end{aligned}
\end{equation}
Here we separate the joint probability into two parts: 
\begin{itemize}
\item $P(y_q=i|X_q,\Theta)$ models the 
estimated distribution of
the `true'
answer $y_q$ from 
the
provided 
reviews.
\item $\prod_{j=1}^{n_q} P(y_{q,j}|y_q=i, X_q, \Theta)$ models
the
probability of a given ground-truth label $y_{q,j}$ as a function of $y_q$.
\end{itemize}
Letting $\alpha_q = P(y_{q,j}=1|y_q=1, X_q, \Theta)$ and $\beta_q = P(y_{q,j}=0|y_q=0, X_q, \Theta)$ 
for all $j\in\mathcal{S}_q$, 
then $\alpha_q$ and $\beta_q$ represent the `sensitivity' (probability of a positive observation if the true
label is positive) and `specificity' (probability of a negative observation
if the
label is negative) 
for
question $q$.
Note
that 
`positive' and `negative' questions may not be symmetric concepts (i.e., different types of questions may be more likely to have yes vs.~no answers). %
Thus we model sensitivity and specificity separately, 
using features from the question text
as prior knowledge. Specifically, we model $\alpha$ and $\beta$ as:
\begin{equation}
\small
\alpha_q = \sigma(\left<\bm{\gamma_1}, \bm{f}_q\right>);\qquad \beta_q = \sigma(\left<\bm{\gamma}_2, \bm{f}_q\right>). \label{eq:alpha_beta}
\end{equation}

Then we have the following joint distributions which are denoted as $a_q$ and $b_q$:
\begin{equation}
\small
\begin{aligned}
a_q := \prod_{j=1}^{n_q} P(y_{q,j}|y_q=1, X_q, \Theta) =& \alpha_q^{n_q^+}(1-\alpha_q)^{n_q^-} \\
b_q := \prod_{j=1}^{n_q} P(y_{q,j}|y_q=0, X_q, \Theta) =& (1-\beta_q)^{n_q^+}\beta_q^{n_q^-}. \label{eq:a_b}
\end{aligned}
\end{equation}

Now based on \eqref{eq:multi}, \eqref{eq:alpha_beta}, and \eqref{eq:a_b}, we can consider maximizing following log-likelihood:
\begin{equation}
\small
\begin{aligned}
    \mathcal{L} =& \log P(\mathcal{Y} | \mathcal{X}, \Theta)
   =\sum_q \log P(y_{q,j}, j=1,\ldots,n_q | X_q)\\
   =& \sum_q \log \left( a_q p_q + b_q (1-p_q)\right),
    \label{eq:log-lik_multi}
\end{aligned}
\end{equation}
where $p_q = P(y_q = 1|X_q, \Theta)$ is modeled based on \eqref{eq:MoE_binary}, \eqref{eq:rel_pred}, \eqref{eq:score_basic_v} and \eqref{eq:score_basic_w}. Here the parameter set is $\Theta = \{\bm{\kappa}, \bm{\eta}, \bm{\mu}, \bm{u}, \bm{\gamma}_1, \bm{\gamma}_2\}$.
\paragraph*{Inference}
In contrast to
\textbf{MoE} and \textbf{KL-MoE},
directly optimizing \eqref{eq:log-lik_multi} is non-trivial. However, we can apply 
the
EM Algorithm \cite{dempster1977maximum} to optimize it by
estimating the
label $y_q$ and the parameters $\Theta$ iteratively. 

By introducing the missing 
labels $\{y_q\}$, we have a complete likelihood expression
\begin{equation}
\small
\mathcal{L}_c= \sum_q \left(y_q \log a_q p_q + (1-y_q)\log b_q (1-p_q)\right). \label{eq:complete-log-lik}
\end{equation}

\begin{itemize}
\item In the \textbf{E-step}, we assume 
that
parameters $\Theta$ are given. Then we take the expectation of $y_q$ in \eqref{eq:complete-log-lik} and we obtain a new objective:
\begin{equation}
\small
\mathbb{E}\mathcal{L}_c= \sum_q \left(t_q \log a_q p_q + (1-t_q)\log b_q (1-p_q)\right),
\end{equation}
where
\begin{small}
\begin{equation*}
t_q = P(y_q = 1 | y_{q,1}, ..., y_{q,n_q}, X_q) = \frac{a_q p_q}{a_q p_q + b_q (1-p_q)}. %
\end{equation*}
\end{small}
\item In the \textbf{M-step}, once $t_q$ is obtained, similar to \textbf{MoE} and \textbf{KL-MoE}, we can apply L-BFGS \cite{nocedal2006numerical} 
to optimize $\mathbb{E}\mathcal{L}_c$ with respect to $\Theta$.
\end{itemize}
These two procedures are repeated until convergence. 

\subsection{Incorporating Subjective Information}
\label{sec:binarySubjective}
\textbf{EM-MoE-S.} 
Subjective information from reviews (and review\emph{ers}) can be included to enhance the performance of both our relevance and prediction functions, including features
such as review helpfulness, reviewer expertise, rating scores and reviewer biases. We can 
incorporate these features into our previous expressions for $v_{q,r}$ and $w_{q,r}$
as follows:
\begin{equation}
\small
\begin{aligned}
v_{q,r} = & \!\overbrace{\left< \bm{\kappa}, \bm{s}(q,r)\right>}^{\text{pairwise similarities}}\! + \!\overbrace{\left< \bm{\eta}, \bm{f}_q \circ \bm{f}_r \right>}^{\text{term-to-term similarity}}\! + \!\overbrace{\vphantom{\left<gh\right>} \left<\bm{g}, \bm{h}_r\right>}^{\text{\textcolor{blue}{review's helpfulness}}} \!+\!\overbrace{\vphantom{\left<gh\right>} e_{u_r}}^{\text{\textcolor{blue}{reviewer's expertise}}}\\[2mm]
w_{q,r} = & (\!\!\!\!\underbrace{\left< \bm{\mu}, \bm{f}_q \circ \bm{f}_r \right>}_{\text{interaction bet.~q.~\&r.~text}} + \!\!\underbrace{\left<\bm{\xi}, \bm{f}_r\right>}_{\text{prediction from r.~text}}\!\!\!\!\!\!\!\!)\times (1 + \underbrace{\vphantom{\left<gh\right>}c\cdot rt_r}_{\text{\textcolor{blue}{rating score}}} + \underbrace{\vphantom{\left<gh\right>}b_{u_r}}_{\text{\textcolor{blue}{reviewer's bias}}}\!\!\!\!\!\!\!).\\
\label{eq:score_f_binary_subj}
\end{aligned}
\end{equation}
As
shown
in \reffig{figure:subjectiveQA}, here $rt_r$ is the star rating score and 
$\bm{h}_r=(h_r^{(1)}, h_r^{(2)})^T$ represents the helpfulness features of review $r$ where $h_r^{(1)}, h_r^{(2)}$ are fractions of users who respectively find or do not find the review helpful.
$e_{u_r}$ and $b_{u_r}$ are parameters that make up a simple user model; the former captures the overall tendency for a user $u$ to write reviews that are likely to be `relevant,' while the latter captures the tendency of their reviews to support positive responses.
Note that both parameters are latent variables that are automatically estimated when we optimize the likelihood expression above.

\section{Modeling Open-ended Questions}

Although our \textbf{KL-MoE} and \textbf{EM-MoE} frameworks can model 
ambiguity
in
binary questions, and account for simple features encoding subjectivity,
we still need to develop 
methods to account for ambiguity in
open-ended questions. 
Here we are no longer concerned with divergence between yes/no answers, but rather want to model the idea that there is a pool of answers to each question which should be regarded as more valid than alternatives.
As with
binary questions, these open-ended questions may be subjective and multiple answers
often exist in our data.
What is different is that
it is difficult for us to automatically judge 
whether
these answers are consistent or not. Thus we aim to
generate candidate answers 
that %
cover 
the spectrum of ground-truth answers as much as possible.
First we give some more detail about the basic framework with a single open-ended answer, which we described briefly in Section \ref{sec:moeOpenEnded}.
Then we simply extend this framework to include multiple open-ended answers and incorporate subjective information.

\subsection{Basic Framework: Learning with 
a
Single Answer.}
\textbf{s-MoE.}
Our objective
for open-ended 
questions
is to maximize the \textit{Area Under Curve (AUC)}, which 
is
defined as
\begin{equation}
\small
AUC_o = \frac{1}{|Q|}\sum_q AUC(q) = \frac{1}{|Q|}\sum_q(\frac{1}{|\mathcal{\bar{A}}_q|} \sum_{\bar{a}\in\mathcal{\bar{A}}_q} \delta(a_q>\bar{a})).
\end{equation}
where $a_q$ is the 
ground-truth
answer to the question $q$ and $\mathcal{\bar{A}}_q$ is a set of non-answers
(randomly 
sampled
from 
among all answers).
In other words, a good system is one that can correctly determine which answer is the real one.\footnote{Note that in practice, at test time, one would not have a selection of candidate answers to choose from; the purpose of the model in this case is simply to identify which reviews are relevant (by using the answers at \emph{training} time), rather than to answer the question directly.}

In practice,
we maximize a smooth objective 
to approximate this measure
in the form of the log-likelihood:
\begin{equation}
\small
\mathcal{L} = \sum_q \sum_{\bar{a}\in\mathcal{\bar{A}}_q}  \log p_{q,a_q>\bar{a}}.
\end{equation}
Here $p_{q,a_q>\bar{a}} = P(a_q>\bar{a}|X_q)$ is 
as defined in
\eqref{eq:MoE_open}. The `relevance' term in $p_{q,a_q>\bar{a}}$ is the same as 
for binary questions while the `prediction' term is defined as
\begin{equation}
\small
p_{q,a_q>\bar{a}|r} = \sigma(w_{a_q>\bar{a}|r}).
\end{equation}
As before,
$w_{a_q>\bar{a}|r}$ can 
be modeled 
in terms of
answer and review text.
Letting
$\bm{f}_{a_q}$ and $\bm{f}_{\bar{a}}$ denote the text features of the answer $a_q$ and the non-answer $\bar{a}$ respectively. Then we have
\begin{equation}
\small
w_{a_q>\bar{a}} = w_{a_q, r} - w_{\bar{a}, r} 
= \overbrace{\left< \bm{\mu}, (\bm{f}_{a_q} - \bm{f}_{\bar{a}}) \circ \bm{f}_r \right>}^{\mathclap{\text{interaction between ans.~difference \& review text}}}. \label{eq:score_open_w}
\end{equation}
$\bm{f}_{a_q} - \bm{f}_{\bar{a}}$ represents the difference between the answers $a_q$ and $\bar{a}$,
so that
\eqref{eq:score_open_w}
models which of the answers $a_q$ or $\bar{a}$ is \emph{more supported} by review $r$.

\subsection{Incorporating Subjective Information with Multiple Answers.}
\textbf{m-MoE.} 
The 
previous AUC measure can be straightforwardly extended to be compatible with multiple answers. If multiple answers exist for a question $q$, then our target is to maximize 
the
following AUC measure:
\begin{equation}
\small
AUC_o = \frac{1}{|Q|}\sum_q(\frac{1}{|\mathcal{A}_q||\mathcal{\bar{A}}_q|} \sum_{a\in\mathcal{A}_q}\sum_{\bar{a}\in\mathcal{\bar{A}}_q} \delta(a>\bar{a})).\label{eq:AUC_open}
\end{equation}
where $\mathcal{A}_q$ denotes the \emph{set} of answers to question $q$ and $\mathcal{\bar{A}}_q$
is defined as before.

Similarly, 
we maximize 
the
following log-likelihood loss function to approximately 
optimize the AUC:
\begin{equation}
\small
\mathcal{L} = \sum_q \frac{1}{|\mathcal{A}_q|}\sum_{a\in\mathcal{A}_q} \sum_{\bar{a}\in\mathcal{\bar{A}}_q}  \log p_{q,a>\bar{a}}. \label{eq:log-lik-open-all}
\end{equation}

\subsection{Incorporating Additional Information from Reviews}
(\textbf{m-MoE-S.}) Similar to binary questions, we can incorporate more subjective features into $v_{q,r}$ and $w_{a>\bar{a},r}$. Basically, $v_{q,r}$ can be kept the same as in \eqref{eq:score_f_binary_subj}. For $w_{a_q>\bar{a},r}$, we have
\begin{equation}
\small
w_{a_q>\bar{a},r} = \overbrace{\underbrace{\left< \bm{\mu}, (\bm{f}_{a_q} - \bm{f}_{\bar{a}}) \circ \bm{f}_r \right>}_{\mathclap{\text{interaction b.w.~ans.~difference \&r. text}}}}^{\text{which answer the review favors}} \times \overbrace{(1 + \underbrace{c\cdot rt_r}_{\text{\textcolor{blue}{rating score}}} + \underbrace{b_{u_r}}_{\text{\textcolor{blue}{reviewer's bias}}})}^{\text{how supportive based on the review}}. \label{eq:score_f_open_subj} 
\end{equation}
The left part of this formula is the same as in \eqref{eq:score_open_w} which models which 
of the answers the review favors.
The right part of this formula is an amplifier which models how supportive the review $r$ is based on its subjective information.

\section{Dataset and Exploratory Analysis}
\label{sec:dataset}

In \cite{mcauley2016addressing}, the authors collected Q/A data from \textit{Amazon.com}, 
including a single answer (the top-voted)
for each question. We collected all the related urls in this dataset and 
further
crawled
all 
available answers to each question (duplicates were discarded, as were questions that have been removed from \emph{Amazon} since the original dataset was collected). For each product we also have its related reviews.
Ultimately
we obtained around 808 thousand questions with 3 million answers on 135 thousand products in 8 large categories. For these products, we have 11 million reviews in total. Detailed information 
is shown
in \reftable{table:basicStats}.
\begin{table}
\begin{center}
\setlength{\tabcolsep}{3pt}
\footnotesize
\begin{tabular}{lrrrr}
\toprule
\textbf{Category} & \textbf{\#products} & \textbf{\#questions} & \textbf{\#answers} &\textbf{ \#reviews} \\
\midrule
Automotive & 10,578 & 59,449 & 233,784 & 325,523 \\
Patio, Lawn \& Garden & 7,909 & 47,589 & 193,780 & 450,880 \\
Tools \& Home Improv. & 13,315 & 81,634 & 327,597 & 751,251 \\
Sports \& Outdoors & 19,102 & 114,523 & 444,900 & 988,831 \\
Health \& Personal Care & 10,766 & 63,985 & 255,209 & 1,154,315 \\
Cell Phones & 10,320 & 60,791 & 237,220 & 1,353,441 \\
Home \& Kitchen & 24,329 & 148,773 & 611,335 & 2,007,847 \\
Electronics & 38,959 & 231,556 & 867,921 & 4,134,100 \\
\midrule
Total & 135,278 & 808,300 & 3,171,746 & 11,166,188 \\
\bottomrule
\setlength{\tabcolsep}{6pt}
\end{tabular}
\end{center}
\caption{Basic statistics of our Amazon dataset.}\label{table:basicStats}
\end{table}
\begin{figure}
    \centering
    \begin{subfigure}[b]{0.235\textwidth}
        \centering
        \includegraphics[width=\textwidth]{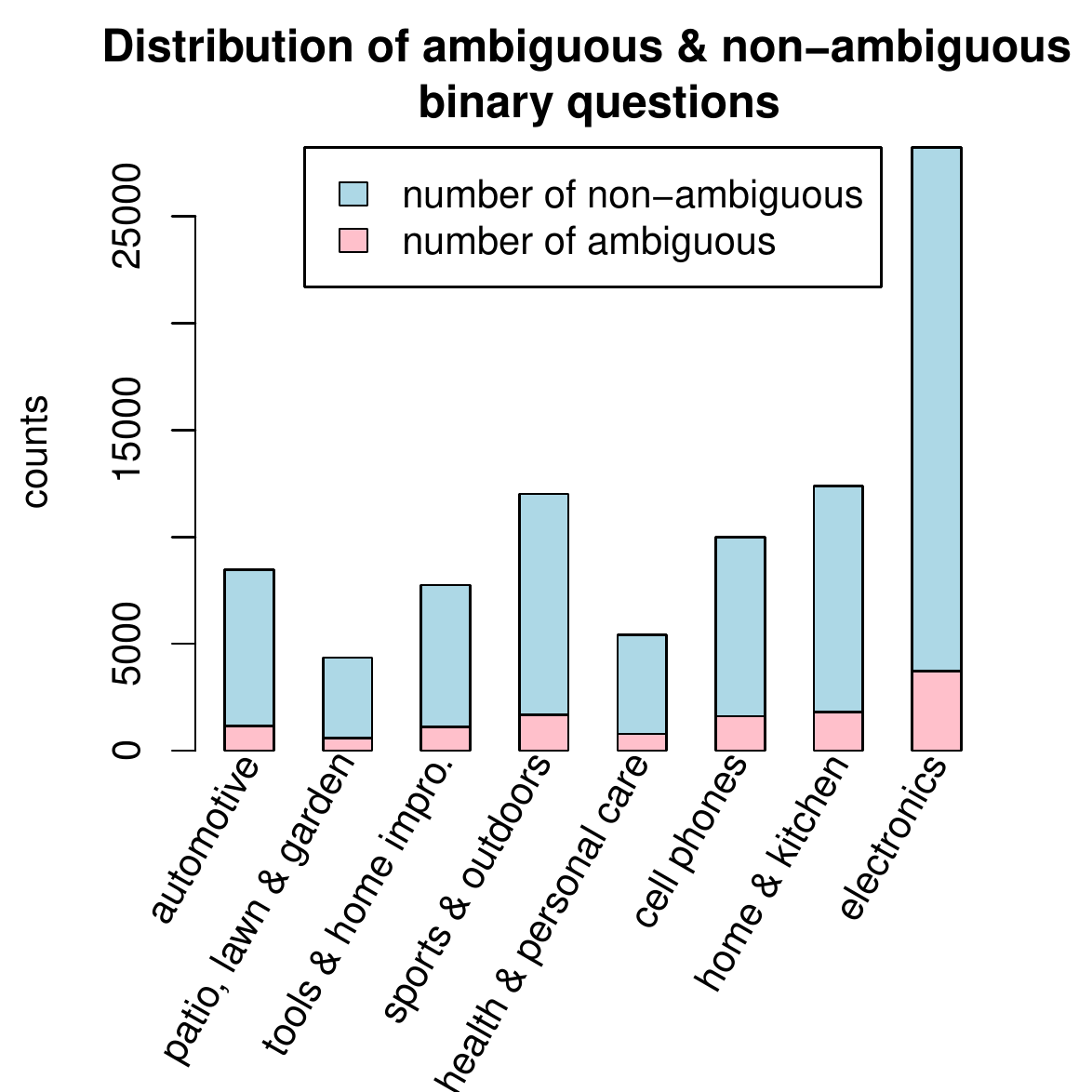}
        \caption{Distribution of ambiguous and non-ambiguous binary questions.}\label{figure:binaryDist}
    \end{subfigure}%
    \hfill
        \begin{subfigure}[b]{0.235\textwidth}
        \centering
        \includegraphics[width=\textwidth]{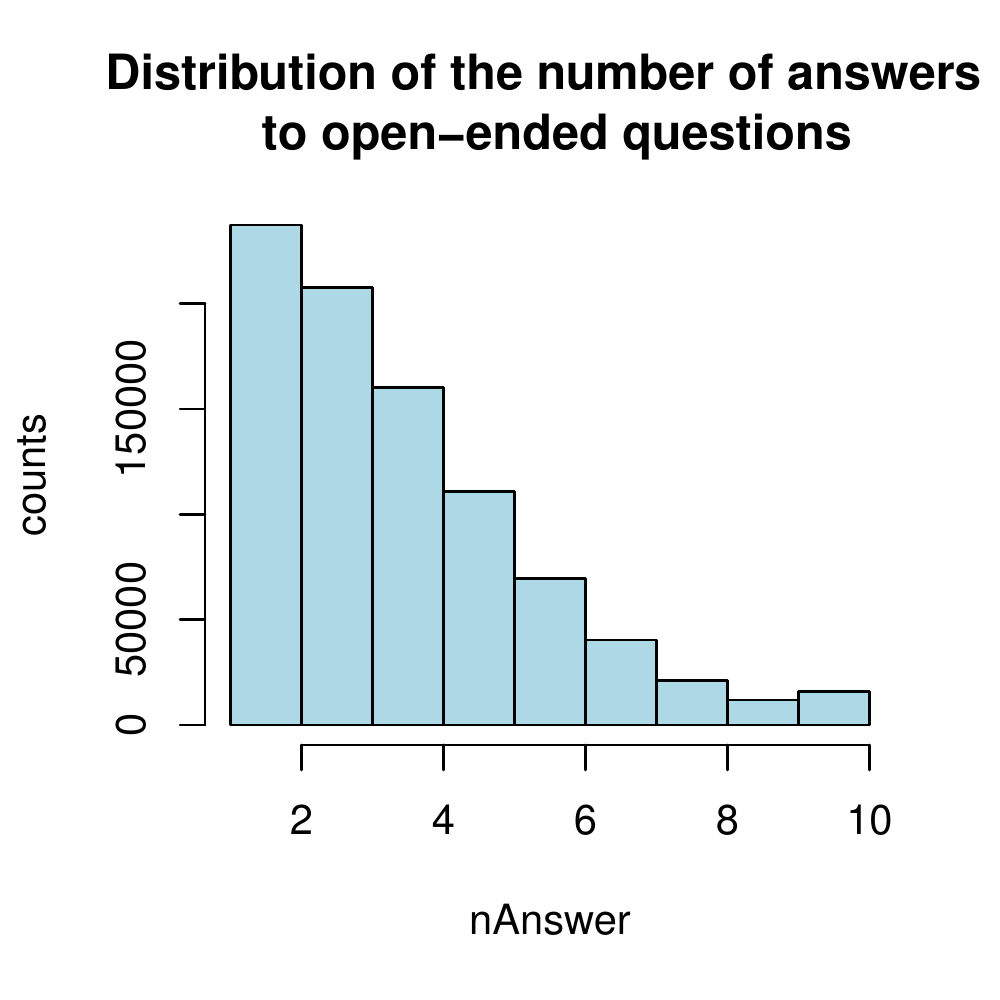}
        \caption{Histogram of numbers of answers for all open-ended questions}\label{figure:openDist}
    \end{subfigure}%
    \caption{Distribution of the dataset.}
\end{figure}
In practice,
we split review paragraphs into sentences, such that each sentence is treated as a single `expert' in our MoE framework.
We used the
Stanford CoreNLP \cite{manning-EtAl:2014:P14-5}
library
to split 
reviews 
into sentences, handle word tokenization, etc.%

\subsection{Obtaining Ground-Truth labels for Binary Questions}
In the dataset from \cite{mcauley2016addressing}, 
one thousand questions
have been
manually labeled as `binary' or `open-ended.' For binary questions, 
a
positive or negative label is provided for each answer. 
We used these labels as seeds to train simple classifiers to identify binary questions with positive and negative answers.
As in \cite{mcauley2016addressing}, we applied 
an
approach developed by \textit{Google} \cite{he2011summarization} to determine whether a question is binary, 
using 
a series 
of simple
grammatical rules.
Among our labeled data, 
this approach achieved 97\% precision and 82\% recall in this manually labeled dataset.\footnote{Note that we are happy to sacrifice some recall for the sake of precision, as low recall simply means discarding some instances from our dataset, as opposed to training on incorrectly labeled instances.}

Following this,
we developed a 
simple
logistic regression model to label observed answers to these binary questions. The features we applied are the frequency of each unigram plus whether the first word is `yes' or whether it is `no' (as is often the case in practice). Notice that since we 
want
to study 
ambiguity 
that arises due to the
question itself 
rather than due to any error in our machine labels,
we need to 
ensure
that 
the
binary labels obtained from 
this
logistic model are as accurate as possible. 
Thus again
we sacrifice some recall %
and keep only those answers about which the regressor is most confident (here we kept the top 50\% of most confident predictions).
This gave us zero error on the held-out manually labeled data from \cite{mcauley2016addressing}. 
Ultimately we
obtained 88,559 questions with 197,210 highly confident binary labels of which around 65\% are positive. In our experiments, 
two thirds
of these questions and associated labels are involved in training and 
the rest
are used for evaluation.

\subsection{Exploratory Analysis}

Having constructed classifiers to label our training data with high confidence,
we next want
to determine whether there 
really
are conflicts between multiple 
answers for 
binary questions.
The distribution of ambiguous (i.e., both yes and no answers)
versus
non-ambiguous binary questions is
shown
in \reffig{figure:binaryDist}. From this figure, we notice that we do have a portion of binary questions that 
can be
confidently classified as `ambiguous.' A real-world example of 
such a question
is 
shown
in \reffig{figure:subjectiveQA}. 
One might expect that the answer to such a question would be an unambiguous `yes' or `no,' since it is a (seemingly objective) question about compatibility. However the answers prove to be inconsistent since different users focus on different aspects of the product.
Thus %
even seemingly objective questions
question can 
prove to be
`ambiguous,' 
demonstrating the need for a model that handles such conflicting evidence.
Ideally a system to address such a query 
would
retrieve relevant reviews 
covering
a variety of angles and in this case provide an overall
neutral
prediction. 

Ultimately, 
around
14\% of the questions in our dataset are ambiguous (i.e., multiple binary labels are inconsistent).
Distributions of ambiguous/non-ambiguous questions are plotted in \reffig{figure:binaryDist}. 
Even though we filtered our dataset to include only answers with high-confidence labels (i.e., clear `yes' vs.~`no' answers),
there is still a significant number of questions with conflicting labels, 
which indicates that modeling ambiguity is necessary for opinion QA systems.
\section{Experiments}
We evaluate our proposed methods for binary questions and open-ended questions on a large dataset composed of questions, answers and reviews from \textit{Amazon}. For binary questions, we evaluate the model's ability to make proper predictions. For open-ended questions, we evaluate the model's ability to distinguish `true' answers from alternatives. Since 
our main goal is to
address
ambiguity and subjectivity,
we 
focus on evaluating our model's ability
to exploit
multiple labels/answers, and
the effect of features derived from subjective information.

\subsection{Binary Questions}
\subsubsection{Evaluation Methodology}

For yes/no questions, our target is to evaluate whether
our
model can predict their `true' labels 
correctly.
Since multiple labels are collected for a single question, 
and since we are comparing against methods capable of predicting only a single label, 
it is difficult to evaluate which system's predictions are most `correct' in the event of a conflict.
Thus for evaluation
we build two test sets consisting of decreasingly ambiguous questions. 
Our hope then is that by modeling ambiguity and personalization during {training}, our system will be more reliable even for unambiguous questions. We build two evaluation sets as follows:
\begin{itemize}
\item \textbf{Silver Standard Ground-truth.} 
Here we simply regard the majority vote among ambiguous answers as the `true' label (questions with ties are discarded).
\item \textbf{Gold Standard Ground-truth.} More aggressively, here we 
ignore all 
questions with conflicting labels. For the remaining 
questions, we have consistent labels that we 
regard 
as ground-truth. 
\end{itemize}
Notice that all the questions and labels (ambiguous or otherwise) in the training set are involved in the training procedure of \textbf{KL-MoE}, \textbf{EM-MoE} and \textbf{EM-MoE-S}. We only attempt to resolve ambiguity when building our test set for evaluation.
Naturally, it is not possible to address all questions using the content of product reviews.
Thus
we are more interested 
in the probability that the model will rank a random positive instance higher than a random negative one.
We adopt the standard AUC measure, which for binary predictions is defined as:
\begin{equation}
\small
\mathit{AUC}_b = \int_{-\infty}^{\infty} \mathit{TPR}(t)\ d(\mathit{FPR}(t)),
\end{equation}
where $t$ is a threshold between 0 and 1. 
Suppose the 
label for question $q$ is $y_q$ and the predicted probability of being positive from a particular model is $\hat{p}_q$. Then we have
\begin{equation*}
\small
\begin{aligned}
TPR(t) = \frac{\sum_q \bm{1}_{\hat{p}_q\ge t, y_q=1}}{\sum_q \bm{1}_{y_q=1}};\qquad FPR(t) = \frac{\sum_q\bm{1}_{\hat{p}_q\ge t, y_q=0}}{\bm{1}_{y_q=0}}.
\end{aligned}
\end{equation*}
Note
that this 
is different from
the AUC in equation
\eqref{eq:AUC_open}, which is in the context of open-ended 
questions.
Note that a na\"ive classifier (random predictions, random confidence ranks) has an AUC of 0.5.

\subsubsection{Baselines}

We compare the 
performance
of the following methods:
\begin{itemize}
\item \textbf{MoE.} 
This is a state-of-the-art method for opinion QA from \cite{mcauley2016addressing}. This is the model described in Section \ref{sec:moeBackground}. Here only a single label (the top-voted) is used for training, and text features from the reviews are included.
\item \textbf{KL-MoE.} This is a straightforward approach to include multiple labels 
by replacing a single label
$y_q$ by the
ratio of positive vs.~negative answers ($r_q = n_q^+/(n_q^+ + n_q^-)$ in \eqref{eq:log-lik} (see Sec.~\ref{sec:klem}).
\item \textbf{EM-MoE.} To include all the labels instead of just a 
ratio,
we use an EM-like approach to update our estimates of noisy labels and parameters iteratively.
Here question text features are used as prior knowledge to model the `sensitivity' and `specificity' regarding a question. 
\item \textbf{EM-MoE-S.} 
Note that the above models only make use of features from reviews, and are designed to measure the performance improvements that can be achieved by harnessing multiple labels.
For our final method, we include other subjective information into our model, such as user bieses, rating features, etc. (see Sec.~\ref{sec:binarySubjective}).
\end{itemize}
Ultimately the above baselines are intended to demonstrate: (a) the performance of the existing state-of-the-art (\textbf{MoE}); (b) the improvement from leveraging conflicting labels during training (\textbf{KL-MoE} and \textbf{EM-MoE}); and (c) the improvement from incorporating additional subjective information in the data (\textbf{EM-MoE-S}).

\subsubsection{Results and Discussion} 
Results of the above methods in terms of the AUC 
are shown in
\reftable{table:resultBinary}. 
We notice that while \textbf{KL-MoE} 
is
not able to improve 
upon
\textbf{MoE} for all categories,
\textbf{EM-MoE} and \textbf{EM-MoE-S} 
yield
consistent improvements
in all cases.\footnote{Improvements in accuracy over \textbf{MoE} are statistically significant at the 1\% level or better.}
This improvement is relatively 
large
for some large categories, such as \emph{Cell Phones \& Accessories} and \emph{Electronics}. 
Incorporating subjective features (\textbf{EM-MoE-S}) seems to help most for large categories, indicating that it is useful when enough training data is available to make the additional parameters affordable.

When 
modeling 
ambiguity in opinion QA systems, a possible reason for the failure of \textbf{KL-MoE} is that the ratio $r_q$ involved in the objective function may not be a representative label for training. If the observed positive label ratio $r_q$ does not properly reflect the 
`true' distribution, it could %
adversely affect
the optimization procedure. In our EM-like frameworks, i.e.,~\textbf{EM-MoE} and \textbf{EM-MoE-S}, this ratio is replaced by a posterior probability, $t_q$, which is updated iteratively.
These EM-like frameworks are relatively more robust to
data with multiple noisy labels compared with \textbf{KL-MoE}. 

\begin{figure}
\centering
\includegraphics[width=0.5\textwidth]{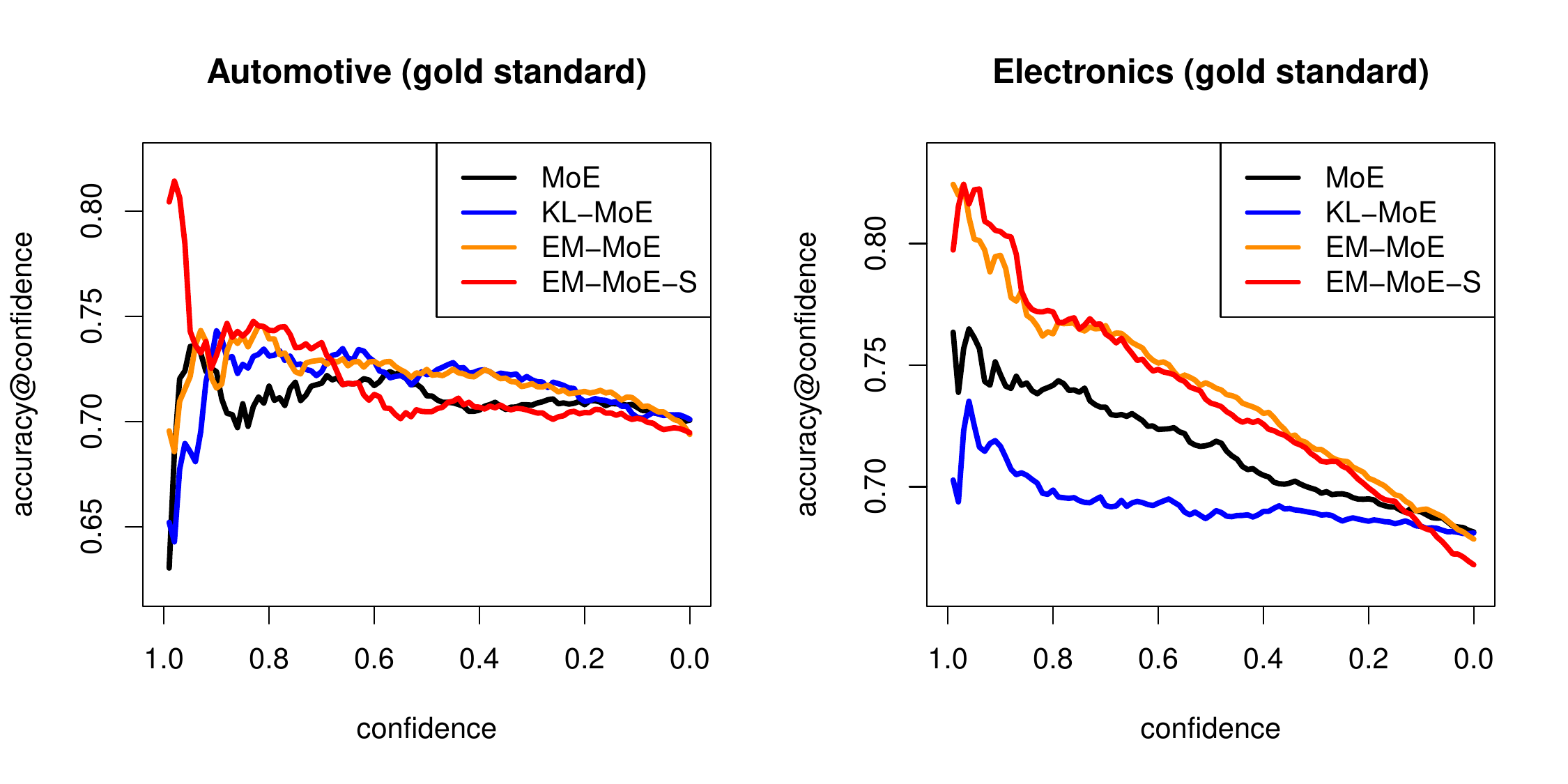}
\caption{Accuracy as a function of confidence on binary questions (Automotive and Electronics categories).}\label{figure:Accur}
\end{figure}

\textbf{EM-MoE-S} 
includes subjective information
related to 
reviews and reviewers. 
Due to the number of parameters involved,
modeling reviewer
expertise and 
bias 
is 
only
useful for 
users who write 
several
reviews, which is indeed a small 
fraction
of 
reviewers. 
Thus in the larger categories these terms appear more useful, once we have enough observations to successfully model them.
Note
that 
the
AUC 
represents the ranking performance on all 
questions. Generally, this value is relatively 
low
in our experiments. 
This is presumably due to the simple fact that many questions cannot be answered based on the evidence in reviews.
Since all of the methods being compared output confidence scores, we are interested in whether competing systems are correct in those instances where they have high confidence.
If $\mathcal{Q}$
denotes the set of all the questions and $\mathcal{Q}_a$ denotes the set of questions 
associated with the first largest $(1-a)|\mathcal{Q}|$ values of $|\hat{p}_q - 0.5|$ (i.e., the most confident about \emph{either} a yes or a no answer),
then we have the following measure for a given confidence threshold $0\le a \le 1$:
\begin{equation}
\small
\mathit{accuracy}@a = \frac{1}{|\mathcal{Q}_a|}\sum_{q\in \mathcal{Q}_a} (\bm{1}_{\hat{p}_q\ge 0.5, y_q=1} + \bm{1}_{\hat{p}_q < 0.5, y_q=0}).
\end{equation}
Recall that 
the
AUC measures the model's ability 
to rank
questions appropriately based on the ground-truth
positive and negative labels. In contrast, 
the $\mathit{accuracy}@a$ instead measures 
the model's ability 
to
correctly predict
labels of those questions with highly confident output ranks. We plot this accuracy score as a function of $a$ for the 
smallest category (\emph{Automotive}) and the 
largest category (\emph{Electronics}) in \reffig{figure:Accur}. We notice that 
the
improvement from modeling ambiguity (\textbf{MoE} vs.~others) is relatively consistent for all 
confidence levels. However, modeling subjective information only seems to improve the performance on 
the most
highly confident instances. For 
a small
category like \emph{Automotive}, since 
there is too little data
to model those 
inactive
reviewers, 
the
\textbf{EM-MoE-S} model 
performs poorly on 
low-confidence instances.

\begin{table}[t]
\begin{center}
\footnotesize
\setlength{\tabcolsep}{3pt}
a) Silver Standard Ground-truth \vspace{0.05in}\\
\begin{tabular}{lcccc}
\toprule
& \textbf{MoE} & \textbf{KL-MoE} & \textbf{EM-MoE} & \textbf{EM-MoE-S} \\
\midrule
Automotive & 0.5226 & 0.5326 & \textbf{0.5354} & 0.5225 \\
Patio Lawn \& Garden & 0.5010 & 0.5184 & \textbf{0.5257} & 0.5173 \\
Tools \& Home Improv.& 0.5514 & 0.5313 & \textbf{0.5690} & 0.5641 \\
Sports \& Outdoors & 0.5536 & 0.5512 & 0.5567 & \textbf{0.5578} \\
Health \& Personal Care & 0.5405 & 0.5157 & 0.5490 & \textbf{0.5588} \\
Cell Phones & 0.5612 & 0.5506 & 0.5936 & \textbf{0.6012} \\
Home \& Kitchen & 0.5087 & 0.5027 & 0.5130 & \textbf{0.5394} \\
Electronics & 0.5525 & 0.5172 & 0.5966 & \textbf{0.6002} \\
\bottomrule
\end{tabular}
\vspace{0.1in}\\

b) Gold Standard Ground-truth\vspace{0.05in}\\
\begin{tabular}{lcccc}
\toprule
& \textbf{MoE} & \textbf{KL-MoE} & \textbf{EM-MoE} & \textbf{EM-MoE-S} \\
\midrule
Automotive & 0.5218 & 0.5363 & \textbf{0.5415} & 0.5285 \\
Patio Lawn \& Garden & 0.5030 & 0.5238 & \textbf{0.5271} & 0.5124 \\
Tools \& Home Improv.& 0.5511 & 0.5280 & \textbf{0.5627} & 0.5547 \\
Sports \& Outdoors & 0.5538 & 0.5491 & 0.5587 & \textbf{0.5628} \\
Health \& Personal Care & 0.5452 & 0.5166 & 0.5530 & \textbf{0.5621} \\
Cell Phones & 0.5661 & 0.5534 & 0.5984 & \textbf{0.6062} \\
Home \& Kitchen & 0.5115 & 0.5052 & 0.5165 & \textbf{0.5382} \\
Electronics & 0.5540 & 0.5171 & 0.5983 & \textbf{0.6046}\\
\bottomrule
\end{tabular}\vspace{0.05in}
\end{center}
\setlength{\tabcolsep}{6pt}
\caption{Results on binary questions where multiple noisy labels are involved.}\label{table:resultBinary}
\end{table}

\subsection{Open-ended Questions}
After our previous procedure to distinguish binary vs.~open-ended questions,
we are left with a total of
698,618 open-ended questions (85\% of all the questions) in 
our
dataset. We plot the distribution of the number of answers provided to each open-ended question in \reffig{figure:openDist} and find that 
the majority
of these questions have more than one 
answer
provided. 

\subsubsection{Evaluation Methodology}
Our goal here is to explore whether using multiple answers at training time can provide us with more accurate results, in terms of the AUC of \eqref{eq:AUC_open}.
In practice,
for each answer $a$, we randomly sample one alternative non-answer 
$\bar{a}$
from the pool of all answers.
Suppose the output probability that answer $a$ to question $q$ is preferred over a non-answer 
$\bar{a}$ is $\hat{p}_{q,a>\bar{a}}$. 
Then the AUC measure 
is defined as
\begin{equation}
\small
\mathit{AUC}_o = \frac{1}{|Q|}\sum_q \frac{1}{|\mathcal{A}_q|}\sum_{a\in\mathcal{A}_q} \bm{1}(\hat{p}_{q,a>\bar{a}}>0.5).
\end{equation}
Note
that although different answers are involved in the training procedures for different models, this evaluation measure is calculated in the same format for the same test data.

\subsubsection{Baselines} We compare 
the performance
of the following methods:
\begin{itemize}
\item \textbf{s-MoE.} This is the method from \cite{mcauley2016addressing}. 
Here only
the top-voted answer 
is included
for training.
\item \textbf{m-MoE.} We include all 
answers for each question in this method and optimize the objective function in \eqref{eq:log-lik-open-all}. Thus we 
evaluate 
whether
training
with
multiple
answers 
improves performance.
\item \textbf{m-MoE-S.} Similarly, we add additional subjective information to our model
in order to
evaluate the contribution of 
subjective features.
\end{itemize}
Again our evaluation is intended to compare (a) the performance of the existing state-of-the-art (\textbf{s-MoE}); (b) the improvement when training with multiple answers (\textbf{m-MoE}); and (c) the impact of including subjective features in the model (\textbf{m-MoE-S}).

\subsubsection{Results and Discussion} Results from \textbf{s-MoE}, \textbf{m-MoE} and \textbf{m-MoE-S} are included in \reftable{table:resultOpen}. 
We find that including
multiple answers in 
our
training procedure 
helps us to obtain slightly better results, while incorporating subjective information 
was not effective here.
A possible reason could be that open-ended questions may not be as polarized as binary questions so that subjective information 
may
not be 
as
good
an indicator 
as compared to the content of the review itself. %

\begin{table}[t]
\footnotesize
\begin{center}
\begin{tabular}{lccc}
\toprule
& \textbf{s-MoE} & \textbf{m-MoE} & \textbf{m-MoE-S} \\
\midrule
Automotive & \textbf{0.8470} & 0.8446 & 0.8459\\
Patio Lawn \& Garden & 0.8640 & \textbf{0.8737} & 0.8673 \\
Tools \& Home Improv. & 0.8676 & \textbf{0.8760} & 0.8680 \\
Sports \& Outdoors & 0.8624  & \textbf{0.8671} & 0.8654 \\
Health \& Personal Care & 0.8697  & \textbf{0.8801} & 0.8218\\
Cell Phones \& Accessories & 0.8326  & \textbf{0.8372} & 0.8232 \\
Home \& Kitchen & 0.8702 & \textbf{0.8746} & 0.8723 \\
Electronics & 0.8481 & \textbf{0.8500} & 0.8480 \\
\bottomrule
\end{tabular}\vspace{0.05in}
\end{center}
\caption{Results on open-ended questions in terms of AUC where multiple answers are involved.}\label{table:resultOpen}
\end{table}

\section{Related Work}
There are several previous studies considering the problem of opinion question answering  \cite{moghaddam2011aqa,yu2012answering,balahur2010going,balahur2010opinion,balahur2009opinion,li2009answering,stoyanov2005multi,mcauley2016addressing}, where questions are subjective and traditional QA approaches may not be as effective as they have been for factual questions.
Yu and Hatzivassiloglou \cite{yu2003towards} first proposed a series approaches to separate opinions and facts and identify the polarities of opinion sentences. Ku \emph{et al.}~\cite{ku2007question} applied a two-layer framework to classify questions and 
estimated
question types and polarities to filter irrelevant sentences. Li \emph{et al.}~\cite{li2009answering} 
proposed a graph-based approach that 
regarded 
sentences as 
nodes 
and weighted edges by sentences similarity;
by constructing such 
a
graph, they could apply an `Opinion PageRank' model and an `Opinion HITS' model to explore different relations. Particularly for product-related opinion QA,
i.e., addressing product-related questions with reviews, an aspect-based approach was proposed where 
aspect-rating data 
were applied \cite{moghaddam2011aqa}.
In Yu \emph{et al.} \cite{yu2012answering}, a new model was developed to generate appropriate
answers for opinion questions by exploiting the hierarchical organization of consumer reviews. Most recently, a supervised learning approach, \emph{MoQA}, was proposed for 
the
product-related opinion QA problem, where a
mixture of experts model was applied and each review was regarded as an expert \cite{mcauley2016addressing}. 

Opinion mining is a broad topic where customer reviews 
are a powerful
resource to explore. A number of opinion mining studies focus on opinion summarization \cite{hu2004mining}, and opinion retrieval and search in review text \cite{liu2006opinion}. In addition, review text can be used to improve recommender systems by modeling different aspects related to customers' opinions \cite{mcauley2013hidden,wang2010latent}. Subjective features and user modeling approaches were frequently applied in these studies, though they were not considered for the opinion QA problem.

The major technique of modeling ambiguity with multiple labels in this study is inspired by approaches for resolving noisy labels in crowdsourcing tasks \cite{raykar2010learning}. 
Notice that the main target of crowdsourcing is to resolve conflicts
from annotators 
and obtain the actual label instead of directly providing accurate predictions from data,
which is different from the setting of answering subjective questions as in our opinion QA problem. In essence, our study can be regarded as a combination of question answering, opinion mining and the idea of learning from crowds.

\section{Conclusion and Future Directions}
In this study, we systematically developed a series of methods to model ambiguity and subjectivity in product-related opinion question answering 
systems.
We 
proposed an EM-like mixture-of-experts framework for binary questions which can successfully incorporate multiple noisy labels and subjective information. Results indicate that this kind of 
framework
consistently 
outperforms
traditional 
frameworks that train using only
a
single label. For open-ended questions, 
we similarly found
that including multiple answers during training improves the ability of the model to identify correct answers at test time.

\ \\
\small
\xhdr{Acknowledgments.} This work is supported by NSF-IIS-1636879, and donations from Adobe, Symantec, and NVIDIA.

\newcommand{\BIBdecl}{\setlength{\itemsep}{0 em}}
\bibliographystyle{IEEEtran}
\bibliography{subjectiveQA}

\end{document}